\documentclass[a4paper,10pt,twoside]{cpc-hepnp}

\usepackage{multicol}
\usepackage{graphicx}
\usepackage{booktabs}
\usepackage{amssymb,bm,mathrsfs,bbm,amscd}
\usepackage[tbtags]{amsmath}
\usepackage{lastpage}
\pdfoutput=1

\begin{document}

\fancyhead[co]{\footnotesize Lu Bo~ et al: A linear calibration method on DNL error for energy spectrum}

\footnotetext[0]{Received 14 March 2009}

\title{A linear calibration method on DNL error for \\energy spectrum\thanks{Supported by HXMT Project}}

\author{%
\quad LU Bo(½²¨)$^{1;1)}$\email{lubo@ihep.ac.cn}%
\quad FU Yan-Hong(¸¶ÑÞºì)$^{1,2}$%
\quad CHEN Yong(³ÂÓÂ)$^{1}$%
\quad CUI Wei-Wei(´Þέέ)$^{1}$%
\and
\quad WANG Yu-Sa(ÍõÓÚØí)$^{1}$%
\quad XU Yu-Peng(ÐìÓñÅó)$^{1}$%
\quad YIN Guo-He(Òü¹úºÍ)$^{1}$%
\quad YANG Yan-Ji(ÑîÑåÙ¥)$^{1,3}$%
\and
\quad ZHANG Yi(ÕÅÒÕ)$^{1}$%
\quad HUO Jia(»ô¼Î)$^{1}$%
\quad LI Wei(Àîì¿)$^{1}$%
\quad CHEN Tian-Xiang(³ÂÌïÏé)$^{1}$%
\and
\quad HU Wei(ºúμ)$^{1}$%
\quad HAN Da-Wei(º«´óì¿)$^{1}$%
\quad LI Mao-Shun(Àîï˳)$^{1}$%
\quad WANG Juan(Íõ¾ê)$^{1}$%
\and
\quad WANG Yu(ÍõÓî)$^{1,4}$%
\quad LIU Xiao-Yan(ÁõÏþÑà)$^{1,3}$%
\quad ZHU Yue(Öì«h)$^{1}$%
\quad ZHAO Zhong-Yi(ÕÔÖÙÒã)$^{1,2}$%
\and
\quad ZHANG Ya(ÕÅæ«)$^{1,2}$%
}
\maketitle

\address{%
$^1$ Institute of High Energy Physics, Chinese Academy of Sciences, Beijing 100049, China\\
$^2$ School of Physical Science and Technology,Yunnan University, Cuihu North Road 2, Kunming 650091, China\\
$^3$ College of Physics, Jilin University, No.2519, Jiefang Road, Changchun 130023, China\\
$^4$ School of Engineering and Technology, China University of Geosciences, Beijing 100083, China\\
}

\begin{abstract}
A calibration method aimed for the differential nonlinearity (DNL) error of the Low Energy X-ray Instrument (LE) onboard the Hard X-ray Modulation Telescope (HXMT) is presented, which is independent with electronic systems used as testing platform and is only determined by the analog-to-digital converter (ADC) itself. Exploring this method, ADCs that are used within the flight model phase of HXMT-LE can be calibrated individually and independently by a non-destructive and low-cost way, greatly alleviating the complexity of the problem. As a result, the performance of the energy spectrum can be significantly improved, further more, noise reduced and resolution enhanced.
\end{abstract}



\begin{multicols}{2}
\section{Introduction}
As one of the main payloads of HXMT\cite{lab1}, the HXMT-LE\cite{lab2} functions to cover a mostly concerned soft X-ray energy band ranging from 1keV to 15keV using a large field of view, while its time resolution is no more than 1ms. To realized this function, several novel Swept Charge Devices (SCDs)\cite{lab3} with a total area as large as 384 cm$^2$ is implemented. Also, the HXMT-LE takes advantage of 12 ADCs to sample and quantize the signals from 96-channel analog front-ends.

According to the function of ADCs in the HXMT-LE application, like most of the particle detector instruments and image-acquisition systems, such static performances as DNL error and integral nonlinearity error are more important than its dynamic counterparts.

\section{Problem and source}
The HXMT-LE is currently at flight model phase and busy going through various space environment simulation tests. The energy spectra obtained by the prototype exhibit a common characteristic, regardless of such factors as electronic system, X-ray source, pixel rate, temperature, operating time, etc.. Fig.~\ref{fig1} is a typical energy spectrum with $^{55}$Fe X-ray source, it is prominent that the count for every fourth bin drops, which is smaller than the counts for both of its adjacent bins. This periodic characteristic covers the full scale output range of the ADC and makes the data points more dispersive along the Gaussian fitting curve.
\begin{center}
\includegraphics[width=6cm]{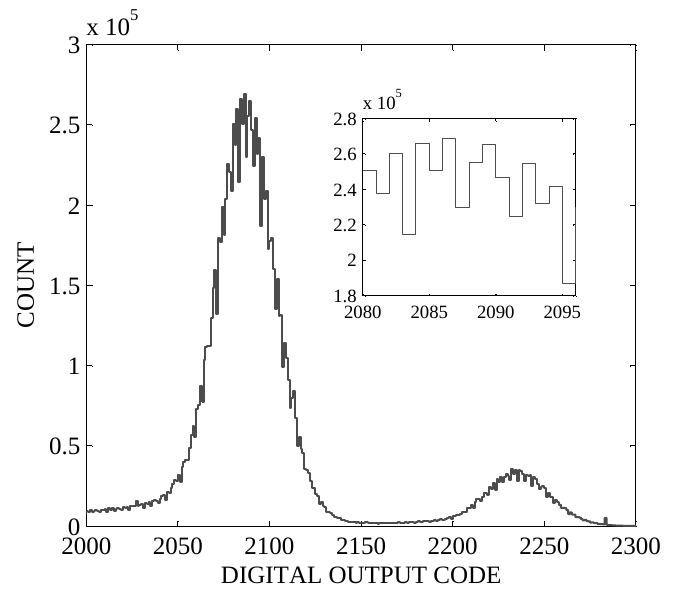}
\figcaption{\label{fig1} Problem of energy spectrum}
\end{center}

A most possible source for the above problem is DNL error. There are a variety of sources responsible for it, among which threshold offset of quantization comparators, periodic spurs dwell on the power lines, nonlinearity of input signal are three most significant sources\cite{lab4}. Even more complicated, some of them come from the ADC itself, while the rest from its electronic environment. Fortunately, for calibration purposes, it is not necessary to distinguish every source of DNL error, rather, the resulting DNL error for every code is sufficient.

\section{Calibration method}
In order to eliminate all of the DNL errors, the goal for the calibration method is to restore the theoretical counts for each code. Based on the problem and its most possible source for the energy spectrum from HXMT-LE, the DNL error for every code is the only useful data that can be resort to. In details, it has to answer two questions for the calibration effect:

1) Is it ADC dependent;

2) Is it electronic system (without ADC) dependent.

The answers will be left to the following section.

\subsection{DNL error and its testing method}
Assume $n$ is the output code, $N$ is the resolution of the target ADC, DNL error is defined as follows\cite{lab5},
\begin{equation}
\label{eq1}
DNL(n)=\frac{V(n)-V(n-1)}{V_{LSB}}-1,n\in[1,2^N-1]
\end{equation}
where $V(n)$ is the corresponding input analog level for the output code $n$ and $V_{LSB}$ is the input analog level that a least significant bit (LSB) represents.  Therefore, DNL errors of two arbitrary adjacent digital codes are intrinsic relevant by definition, lending $DNL(n)$ a dependent variable from the views of statistics. Fortunately, in the following analysis, an independent-event approximation can be reached without sacrificing its accuracy.

Since the dominance in testing the DNL error for medium- and high-speed ADC, as the case in HXMT-LE, the sinusoidal code density method (SCDM) \cite{lab4} is explored in our testing. Taking advantage of DS360\cite{lab6} as the high-quality sinusoidal signal generator and well-designed signal-fed strategy, the testing procedure is facilitated and the liability of the output codes guaranteed.

Assume the theoretical and practical counts for code $n$ are $H_{T}(n)$ and $H_{P}(n)$ respectively, the corresponding DNL error defined by SCDM is\cite{lab4}
\begin{equation}
\label{eq2}
DNL(n)=\frac{H_{P}(n)}{H_{T}(n)}-1
\end{equation}
Under condition of large amount of total counts, Eq.~(\ref{eq2}) makes the $DNL(n)$ an independent variable, which provides a good and reliable approximation to the complicated issue.

As a merit of the target testing method, $H_{T}(n)$ can be easily calculated with the help of spectrum density function of the bathtub curve, a characteristic of sinusoidal wave. Given the total counts ($M_{T}$), and the spectrum density for code $n$ ($P(n)$), $H_{T}(n)$ is calculated as
\begin{equation}
\label{eq3}
H_{T}(n)=P(n) M_{T}
\end{equation}
From this point of view, $H_{T}(n)$ is actually the counts generated by an ideal ADC without DNL error.

It is deserve mention that, the acquisition of DNL error does not require low-temperature environment as the case to obtain the energy spectrum data (ESdata). By contrast, it can be done merely under room temperature, lending this method a good trade-off between the complexity and accuracy.

\subsection{Definition of CF}
With the previous analysis, the calibration method of the target problem is quite straight forward. The first-order calibration factor(CF$_{1}$) is defined as
\begin{equation}
\label{eq4}
CF_{1}(n)=\frac{1}{DNL(n)+1}
\end{equation}
Therefore, the first-order theoretical counts $H_{T1}(n)$ is merely the multiplication of $CF_{1}(n)$ and $H_{P}(n)$.

As an important constraint condition for CF, according to the physical principle, the total counts before and after calibration should be conservative. However, due to the non-idealities in testing the DNL, CF$_{1}$ fails to satisfy the total counts conservation condition. Nevertheless, the resulting total counts error is only about $0.15\%$, indicating that CF$_{1}$ is, in fact, still a fairly good approximation. In consequence, based on CF$_{1}$, it is sound to define the second-order calibration factor(CF$_{2}$) as follows
\begin{equation}
\label{eq5}
CF_{2}\left(n\right)= {\sum {H_{P}(n)}}\left({\sum {\frac{H_{P}(n)}{DNL(n)+1}}}\right)^{-1}CF_{1}(n)
\end{equation}

Since CF$_{2}$ is superior to CF$_{1}$, it is used to present CF hereafter.

\section{Experimental results}
\subsection{Experimental setup}
\begin{center}
\tabcaption{ \label{tab1}  Information for CF}
\footnotesize
\begin{tabular*}{80mm}{c@{\extracolsep{\fill}}ccc}
\toprule \hphantom{000000}CF  & \hphantom{000000}ADC & \hphantom{0000}System\\
\hline
\hphantom{000000}CF1-1 & \hphantom{000000}No.1 & \hphantom{0000}No.1 \\
\hphantom{000000}CF2-1 & \hphantom{000000}No.2 & \hphantom{0000}No.1 \\
\hphantom{000000}CF3-1 & \hphantom{000000}No.3 & \hphantom{0000}No.1 \\
\hphantom{000000}CF1-2 & \hphantom{000000}No.1 & \hphantom{0000}No.2 \\
\hphantom{000000}CF2-2 & \hphantom{000000}No.2 & \hphantom{0000}No.2 \\
\bottomrule
\end{tabular*}
\end{center}

\begin{center}
\tabcaption{ \label{tab2}  Information for ESdata}
\footnotesize
\begin{tabular*}{80mm}{c@{\extracolsep{\fill}}ccc}
\toprule \hphantom{0000}ESdata  & \hphantom{0000}ADC & \hphantom{0000}System\\
\hline
\hphantom{0000}ESdata1-1 & \hphantom{0000}No.1 & \hphantom{0000}No.1 \\
\hphantom{0000}ESdata2-1 & \hphantom{0000}No.2 & \hphantom{0000}No.1 \\
\bottomrule
\end{tabular*}
\end{center}

In order to answer the questions raised at the beginning of the previous section, three ADCs and two sets of electronic systems are adopted. CFs for ADCs from both sets of electronic systems are obtained, shown in Table~\ref{tab1}. In addition, several ESdata using $^{55}$Fe X-ray source are also obtained for two ADCs from an arbitrary set of electronic system, among which, only two arbitrary data are listed in Table~\ref{tab2} for the above purpose.
\subsection{Calibration effects}
The calibration effects using all three CFs from the same No.1 electronic system on ESdata1-1 is shown in Fig.~\ref{fig2}. It is clear that CF from a different ADC with that for ESdata fails for a successful calibration. It concludes that the calibration effect is ADC dependent.

\end{multicols}
\ruleup
\begin{center}
\includegraphics[width=12cm]{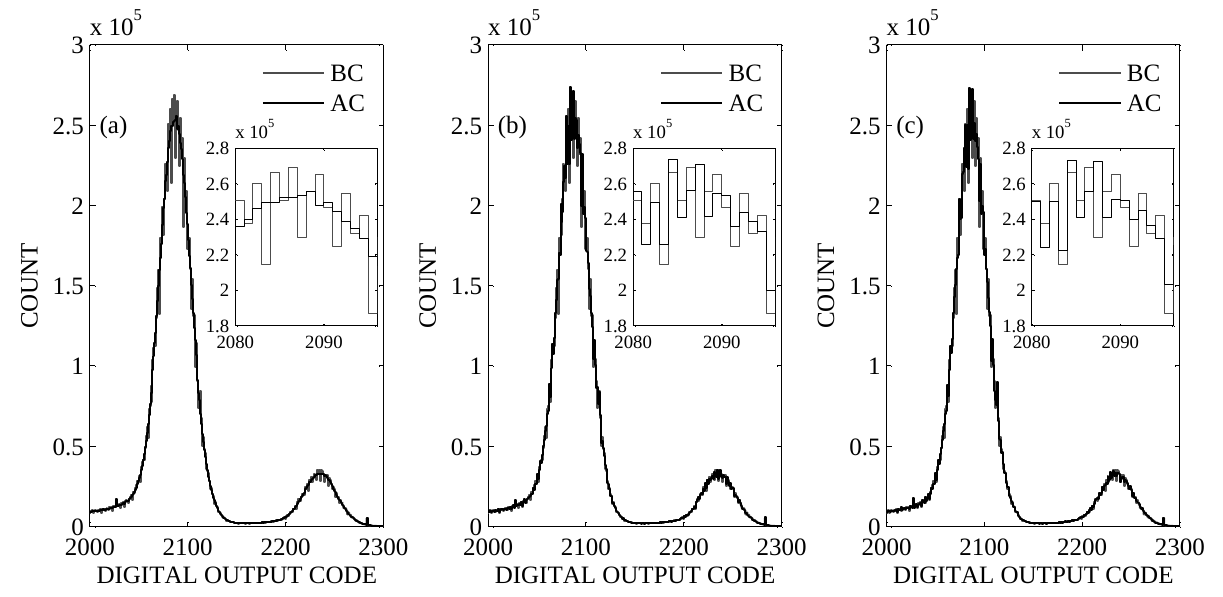}
\figcaption{\label{fig2} Calibration effect on ESdata1-1 using CF1-1, CF2-1 \& CF3-1}
\end{center}
\ruledown

\begin{multicols}{2}
\begin{center}
\includegraphics[width=8cm]{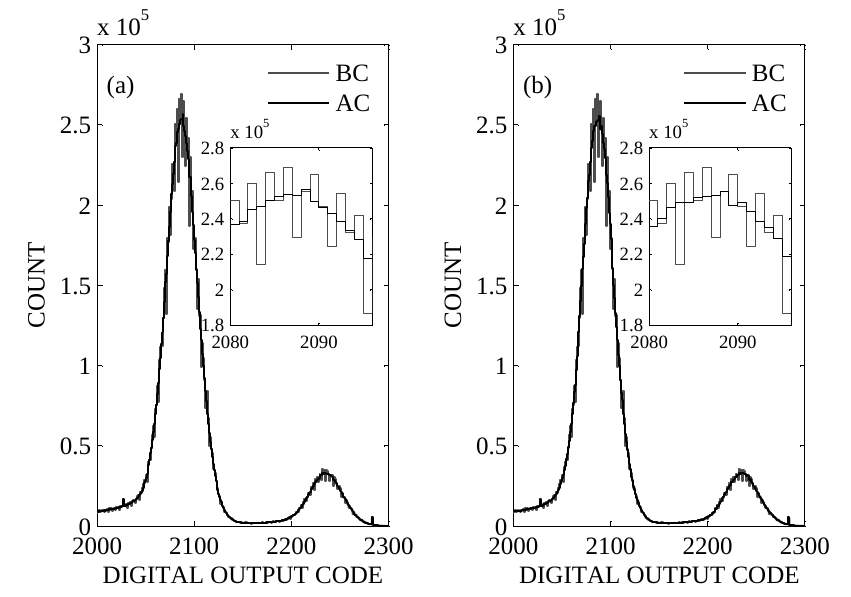}
\figcaption{\label{fig3} Calibration effect on ESdata1-1 using CF1-1 \& CF1-2}
\end{center}

\begin{center}
\includegraphics[width=8cm]{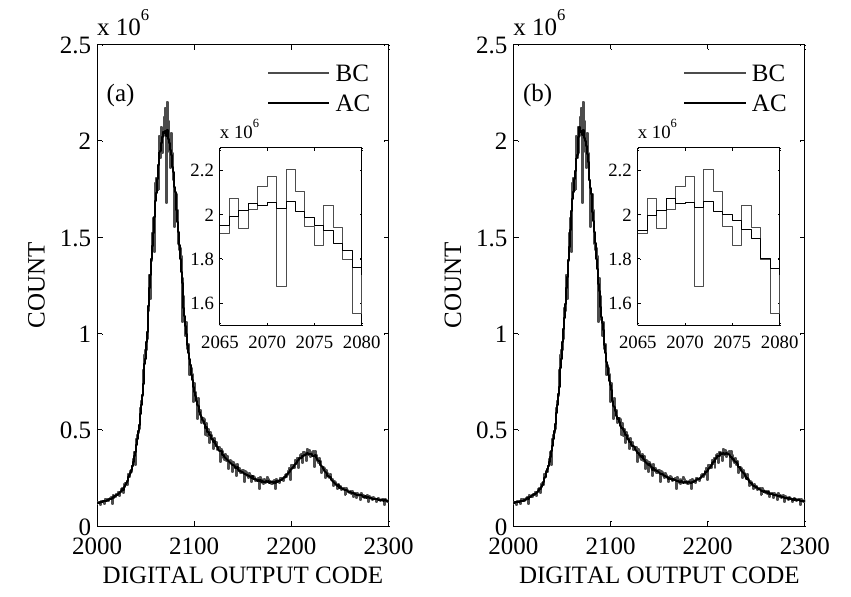}
\figcaption{\label{fig4} Calibration effect on ESdata2-1 using CF2-1 \& CF2-2}
\end{center}

The calibration effects using CFs from No.1 ADC on ESdata1-1 and using CFs from No.2 ADC on ESdata2-1 are shown in Fig.~\ref{fig3} and Fig.~\ref{fig4}, respectively. It is also clear that CF from the same ADC as that for energy spectrum is competent for a successful calibration, despite of the difference between the electronic systems used for obtaining CF and ESdata. It concludes that the calibration effect is electronic system independent.

\subsection{Efficiency and reliability}
In order to verify the efficiency and reliability of the proposed method under conditions with different temperatures and ESdata sizes, the calibration effect has been investigated over a range of temperature from 203K to 223K on the ESdata obtained from the combination of No.1 ADC and No.1 electronic system. In addition, the calibration effects for different data sizes at an arbitrary temperature point, in the vicinity of 213K in this case, have been compared, as shown in Fig.~\ref{fig5}.

It is evident that, both variations of temperature and ESdata size hardly have negative influence on the calibration effect, thus, the proposed calibration method has a good performance on efficiency and reliability.

\end{multicols}
\ruleup
\begin{center}
\includegraphics[width=12cm]{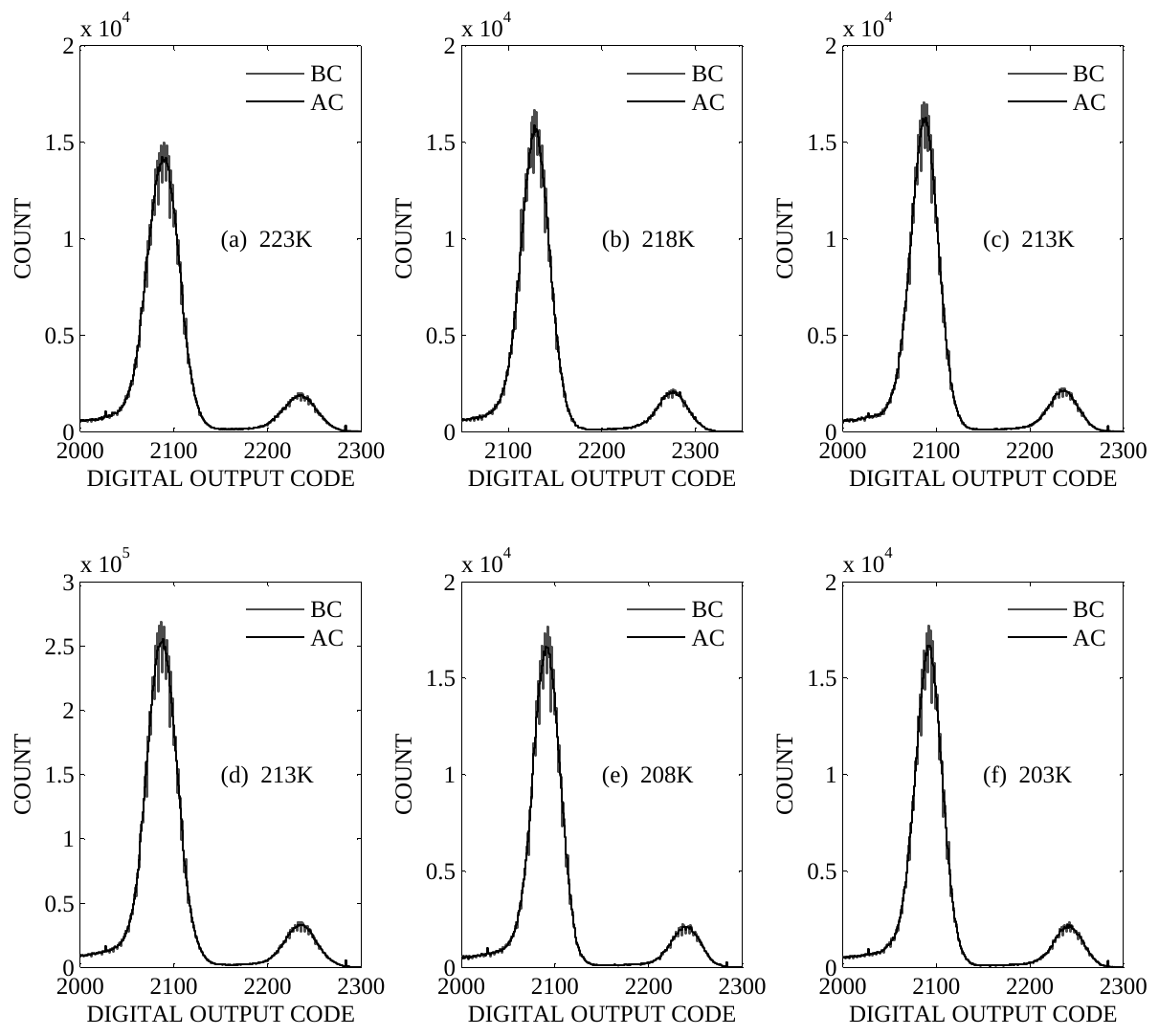}
\figcaption{\label{fig5} Calibration effect on different temperatures and ESdata sizes}
\end{center}
\ruledown
\\
\begin{multicols}{2}

\subsection{Effect on noise and FHWM}
The calibration effects on the details of the energy spectrum, such as noise and full-width-half-magnitude (FWHM), have also been investigated. Using Gaussian fitting method, the noise and FWHM can be easily obtained\cite{lab7}. Table~\ref{tab3} shows some interesting results through 5 ESdata with almost the same data size as shown in Fig.~\ref{fig5}, where the suffix ``BC'' and ``AC'' represent ``before calibration'' and ``after calibration'' respectively. It shows that the calibration method has a positive effect on the performance improvement of both noise and FWHM over a wide temperature range, which is another evidence for the efficiency and reliability of the calibration method.

\end{multicols}
\begin{center}
\tabcaption{ \label{tab3}  Noise and FWHM performances comparison.}
\vspace{-3mm}
\footnotesize
\begin{tabular*}{170mm}{@{\extracolsep{\fill}}ccccccc}
\toprule \hphantom{0000000}ESdata & Temp./$K$  & Noise\_BC/$e^{-}$ & Noise\_AC/$e^{-}$ &  FWHM\_BC/$eV$   & FWHM\_AC/$eV$\\
\hline
\hphantom{0000000}ESdata1-1-1  &223  &13.1 &12.7 &167.9 &167.1\\
\hphantom{0000000}ESdata1-1-2  &218  &8.5  &8.2  &150.2 &145.1\\
\hphantom{0000000}ESdata1-1-3  &213  &7.3  &7.2  &143.7 &141.6\\
\hphantom{0000000}ESdata1-1-4  &208  &6.4  &6.3  &140.8 &139.6\\
\hphantom{0000000}ESdata1-1-5  &203  &6.0  &5.9  &138.8 &137.8\\

\bottomrule
\end{tabular*}%
\end{center}
\begin{multicols}{2}

\section{Conclusions}
Calibration of DNL error for the target application is a challenge work, trading off accuracy with complexity. As a distinguished merit of the proposed calibration method, CFs are obtained under room temperature. With large amount of total counts, the accuracy can be guaranteed despite of the simplicity of the method presented. It has been proved that, CFs for the same ADC from different electronic systems are almost consistent with each other and exhibit excellent calibration effects. It is this important verified conclusion that provides an efficient and reliable way for the calibration of the ADCs used in the flight module of HXMT-LE. More importantly, the present calibration method can be further extended to the similar applications.\\

\end{multicols}

\clearpage

\end{document}